\documentclass[twocolumn,aps,prd,superscriptaddress,amsmath,amssymb,nofootinbib,nobibnotes]{revtex4-2}
\usepackage{graphicx} 
\usepackage{amsfonts, times, amsthm}
\usepackage{bm} 
\usepackage{epstopdf}
\usepackage[linktocpage,colorlinks=true,urlcolor=blue]{hyperref}
\usepackage[dvipsnames]{xcolor}     
\usepackage{natbib}
\usepackage{soul}
\usepackage[utf8x]{inputenc}
\definecolor{darkgreen}{rgb}{0,0.5,0}

\def\p{\partial}

\newcommand{\sss}{\scriptscriptstyle}
\DeclareUnicodeCharacter{2009}{\,} 
\begin{document}
	\title{Traversable wormholes and light rings}

\author{S\'ergio V.~M.~C.~B.~Xavier}
	\email{sergio.xavier@icen.ufpa.br}
	\affiliation{Programa de P\'os-Gradua\c{c}\~{a}o em F\'{\i}sica, Universidade Federal do Par\'a, 66075-110, Bel\'em, Par\'a, Brazil.}
\author{Carlos A.~R.~Herdeiro}
	\email{herdeiro@ua.pt}
	\affiliation{Departamento de Matemática da Universidade de Aveiro and Centre for Research and Development in Mathematics and Applications (CIDMA), Campus de Santiago, 3810-183 Aveiro, Portugal.}
\author{Lu\'is C.~B.~Crispino}
	\email{crispino@ufpa.br}
	\affiliation{Programa de P\'os-Gradua\c{c}\~{a}o em F\'{\i}sica, Universidade Federal do Par\'a, 66075-110, Bel\'em, Par\'a, Brazil.}

\begin{abstract}
Ultracompact objects (UCOs) are horizonless compact objects that present light rings (LRs) - circular photon orbits. As a result, they could be black hole mimickers. Some years ago, Cunha et al. established a theorem stating that, under general assumptions, UCOs formed from smooth, quasi-Minkowski initial data, must have at least a pair of LRs, one of which must be stable. These stable LRs are supposed to trigger a non-linear instability in spacetime, potentially weakening UCOs’ ability to replicate black hole phenomenology. However, this LR theorem does not extend to wormholes, which represent topologically nontrivial spacetimes. We address the wormhole case by proving the following theorem: a stationary, axisymmetric, asymptotically flat, traversable wormhole in 1+3 dimensions, connecting two different asymptotic regions, has at least one standard LR for each rotation sense. Thus, \textit{any} (such) wormhole is an UCO. By filling this gap, our results not only broaden the horizon of knowledge on UCOs but also highlight their potential to closely mimic black hole phenomenology.
	\end{abstract}
	
	\date{\today}
	
	\maketitle

\section{Introduction}\label{sec:int}
Over the last decade, research on the strong gravity regime has entered a golden age. Observational channels -- mainly gravitational waves and shadow images of compact objects \cite{Abbott_etal:2019,Abbott_etal:2021,Abbott_etal:2021gwtc3,M87_1:2019,sgra_1:2022} -- have begun to provide horizon-scale data with remarkable detail. These technological advancements allow rigorous tests of the Kerr hypothesis \cite{Herdeiro:2023}, which states that all astrophysical black holes (BHs) are uniformly described by the Kerr solution, characterized by only two parameters: mass and angular momentum. Underpinning this hypothesis is a series of theorems suggesting that gravitational collapse results in stationary, axisymmetric rotating BHs \cite{Carter:1971, Robinson:1975}, which holds over several mass scales. The Nobel laureate Subrahmanyan Chandrasekhar elegantly captured the essence of such an idea when he termed the realization of this unified description a ``shattering experience” \cite{Chandrasekhar:1975}. 

The challenge of the BH picture is driven by both theoretical puzzles, such as the information loss paradox \cite{Mathur:2009hf}, and phenomenological observations, including gravitational wave detections that leave room for the existence of other compact entities \cite{Maggio_etal:2021, LIGOScientific:2020ufj, CalderonBustillo:2020fyi}. Consequently, it becomes crucial to explore the landscape of horizonless compact alternatives, often referred to as exotic compact objects (ECOs) \cite{Cardoso:2019rvt}. The realm of ECOs is vast, covering entities such as gravastars~\cite{Mazur:2001fv}, boson stars~\cite{Schunck:2003kk, Liebling:2012fv}, fuzzballs~\cite{Mathur:2005zp, Mayerson:2020tpn, Mayerson:2023wck}, and wormholes~\cite{Morris:1988cz, Visser:1995cc, Bambi:2021qfo}. Within this diverse class, certain candidates, dubbed ultracompact objects (UCOs) \cite{Cardoso:2019rvt}, exhibit a very special family of light trajectories known as light rings (LRs). The ability of UCOs to host LRs endows them with the capability to mimic, to some extent, phenomena associated with BH physics, such as shadows and the initial signatures of gravitational wave signals \cite{Cunha:2018acu,Cardoso:2014sna, Cardoso:2016rao}.

LRs represent an extreme form of light bending, manifesting as planar closed light orbits around BHs.~A powerful theorem established their existence in asymptotically flat BH spacetimes,  provided that some reasonable assumptions are satisfied \cite{Cunha_Herdeiro:2020PRL}.  It pioneered a topological approach to the problem, by assigning a topological charge to the critical points of the effective geodesic potential, which correspond to the locations of the LRs. This  method revealed that at least one unstable LR is always present in stationary, axisymmetric and asymptotically flat BH spacetimes. Such result has since been extended to other asymptotic conditions and inspired subsequent research~\cite{Wei:2020rbh,Junior:2021dyw,Junior:2021svb,Wu:2023eml,Zeus_etal:2024,Hod:2017zpi,Ghosh:2021txu,Tavlayan:2022hzl,Cunha:2022nyw,Yin:2023pao,Cunha:2024ajc}. 

In a related work, Cunha et al. showed that UCOs emerging from smooth, quasi-Minkowski initial conditions invariably have at least two LRs, one of which is stable~\cite{Cunha_etal:2017PRL}. The presence of a stable LR introduces a potential for nonlinear spacetime instability, as it can trap and accumulate massless perturbations~\cite{Keir:2014oka}. Recently, the study in Ref.~\cite{Cunha:2022gde} demonstrated that such stable LRs can induce instabilities within a moderate timeframe, leading either to the collapse into BHs or a shift away from the UCO regime. This result challenges the viability of certain UCOs as BH foils, especially those formed from quasi-Minkowski initial data.

While the pioneering results of Refs.~\cite{Cunha_etal:2017PRL,Cunha:2022gde} offer profound insights about LRs in UCOs formed from quasi-Minkowski initial data, the theorem does not include entities with non-trivial topology, such as wormholes. This limitation invites a dedicated examination of wormholes, which stand apart due to their distinctive topological features. Wormholes have been a subject of interest in both physics and science fiction because they can connect different regions of the same universe (intra-universe wormholes) or different universes (inter-universe wormholes), providing a theoretical conception of interstellar or time travel \cite{Morris:1988cz,Morris:1988tu,Frolov:1990si}. Their unique feature of non-trivial topology has made wormholes a compelling object of study. In general relativity (GR), the wormhole throat is usually maintained by a negative pressure that violates the energy conditions imposed on regular baryonic matter~\cite{Morris:1988cz,Visser:1995cc,Visser:1997yn,Kar:2004hc}. There have been various proposals to overcome this issue, such as constraining the energy condition violation to a small region of spacetime~\cite{Visser:1989kg,Visser:2003yf,Poisson:1995sv} or studying solutions from other gravity theories~\cite{Hochberg:1990is, Lobo:2009ip, Kanti:2011jz,Magalhaes:2023har}. Recently, the first asymptotically flat and traversable\footnote{The traversability condition means that massive and massless particles can travel from one asymptotic region to the other without encountering either event horizons or singularities.} wormhole in GR without the need of exotic matter to keep the throat open was found~\cite{Blazquez-Salcedo:2020czn,Konoplya:2021hsm}.

In recent years, the field of wormhole physics has garnered increasing attention~\cite{Shaikh:2019jfr,Simpson:2018tsi,Paul_etal:2020,Lobo:2020kxn,Wielgus:2020uqz,Lima:2020auu,Badia_Eiroa:2020,Maldacena:2020sxe,Guerrero_etal:2021,Volkel_etal:2021,Lobo:2020ffi,Olmo_etal:PLB2022,Guerrero_etal:PRD2022,Halla_Perlick:2022,LimaJunior:2022zvu,	Maldacena:2018gjk,Dias:2023pdx,Huang_etal:2023,Magalhaes:2023xya}. Nevertheless, investigations of the LR structure in spacetimes with nontrivial topology remain remarkably scarce~\cite{Shaikh:2018oul,Bronnikov:2018nub,Tsukamoto:2024pid}. Indeed, a careful analysis of the LR structure of a general wormhole spacetime is lacking in the literature. Wormholes can offer unique gravitational lensing characteristics that could unveil novel observational signatures, such as the existence of multiple critical curves~\cite{Olmo_etal:PLB2022}. Identifying these features is crucial for distinguishing between BHs and wormholes,  thus providing a novel way to test the Kerr hypothesis.

We aim to fill this gap by addressing the question of how many LRs a traversable, axisymmetric, asymptotically flat, inter-universe wormhole geometry can support. Using the topological technique developed in Ref.~\cite{Cunha_Herdeiro:2020PRL}, our analysis spans a broad range of wormhole spacetimes, independent of the gravitational theory. Our findings reveal that a general class of wormholes shares the same LR topological division as BHs, positioning them - at least from this point of view -  as viable candidates for BH mimickers. This result remains true across both static and rotating wormholes, whether symmetric or asymmetric relative to the throat. Furthermore, for the symmetric case, we demonstrate that a LR invariably exists at the throat itself. 

The remainder of this paper is organized as follows. In Sec.~\ref{sec:1}, we introduce the asymptotically flat wormhole spacetime analyzed in this work, in a coordinate system appropriate to describe both sides of the throat. In Sec.~\ref{sec:2}, we discuss the definition of LR in the potential framework and how to assign a topological charge from it. We show that the topological charge of traversable wormhole spacetime, under the conditions assumed in this work, has the same value as asymptotically flat BHs, implying the existence of at least one unstable LR. In Sec.~\ref{sec:3}, we show that wormholes symmetric with respect to the throat will always have a LR at the throat location. We present our final remarks in Sec.~\ref{sec:remarks}. Throughout the paper, we use natural units ($c=G=\hbar=1$) and the metric signature ($-,+,+,+$).

\section{Wormholes spacetime}\label{sec:1}
	
We are interested in stationary, axisymmetric, asymptotically flat spacetimes. These geometries are characterized by having two Killing vectors, denoted as \(\xi\) and \(\eta\), which correspond to the properties of stationarity and axisymmetry, respectively. The concept of asymptotic flatness demands that these vectors must commute. As a result of these properties, it becomes possible to select coordinates \( (t, \varphi) \) that are aligned with these Killing vectors, such that \(\xi\) corresponds to \(\partial_t\) and \(\eta\) to \(\partial_\varphi\).

Furthermore, we assume that the spacetime's metric is at least \( C^2 \) smooth and that it has a circular nature. These assumptions lead to the emergence of a discrete symmetry in the spacetime, represented by the transformation \( (t, \varphi) \rightarrow (-t, -\varphi) \). This symmetry is a direct consequence of the spacetime admitting a 2-dimensional subspace that is orthogonal to the vectors \(\partial_t\) and \(\partial_\varphi\).

Since we are interested in the global properties of the spacetime  of a traversable inter-universe wormhole, it is essential to choose a coordinate patch adequate to analyze the entire manifold. Therefore, the coordinate ranges used here are defined as follows:
\begin{equation}
t, \ell \in \mathbb{R}, \quad \theta \in [0, \pi], \quad \varphi \in [0, 2\pi].
\end{equation}
This coordinate system is designed to cover the full extent of the wormhole spacetime.  It stretches from one asymptotic region, where \(\ell = -\infty\) to the opposite asymptotic region at \(\ell = +\infty\). 

According to Ref.~\cite{Halla_Perlick:2022}, a general class of stationary and axisymmetric inter-universe wormholes, under the metric assumptions described previously,  can be given by:
\begin{align}
ds^2 = & - \mathcal{\tilde{N}}(\ell, \theta)^2 dt^2 + \mathcal{B}(\ell, \theta)^2 d\ell^2 \\
& + \mathcal{\tilde{R}}(\ell, \theta)^2 [d\theta^2 + \sin^2\theta (d\varphi - \omega(\ell, \theta) dt)^2].
\end{align}
This metric is a generalization of rotating wormholes described by Teo in Ref.~\cite{Teo:1998dp}.

Given our focus on null geodesics, which are invariant under conformal transformations, we will henceforth use the metric presented below
\begin{align}
ds^2 = & -\mathcal{N}(\ell, \theta)^2 dt^2 + d\ell^2 \nonumber\\
& + \mathcal{R}(\ell, \theta)^2 [d\theta^2 + \sin^2\theta(d\varphi - \omega(\ell, \theta) dt)^2],
\label{intertwormholemetric}
\end{align}
where we have defined
\begin{equation}
\mathcal{N} := \frac{\tilde{\mathcal{N}}}{\mathcal{B}}, \quad \mathcal{R} := \frac{\tilde{\mathcal{R}}}{\mathcal{B}}.
\end{equation}
The functions $\mathcal{N}$ and $\mathcal{R}$ are defined to be strictly positive. Furthermore, the condition for asymptotic flatness prescribes distinct asymptotic behaviors:
\begin{align}
&\mathcal{N} = 1 + \mathcal{O}(1/|\ell|)\label{Nafconditions}, \\
&\mathcal{R} = |\ell| (1 + \mathcal{O}(1/|\ell|)),\label{afconditions} \\
&\omega = \mathcal{O}(1/|\ell|^2).\label{wafconditions}
\end{align}
The asymptotic behavior of these functions, mainly \(\mathcal{R}(\ell)\), will be important for further analysis in this paper.

Our analysis does not assume any reflection symmetry along the radial coordinate ($\ell \to -\ell$) or along the equatorial plane ($\theta \to \pi - \theta$); thus, it is applicable to both symmetric and asymmetric wormholes. Symmetric wormholes exhibit reflective symmetry concerning the throat, which results in identical geometrical structures on either side of the throat. In contrast, asymmetric wormholes do not possess such reflective symmetry, leading to distinct geometrical features on each side of the throat.

\section{Light rings and topological charge}\label{sec:2}

Building on the wormhole spacetime discussed in the preceding section, we now turn to the study of the null geodesic flow, with a particular focus on the characterization of LRs. These circular photon orbits, pivotal to understanding the optical properties and stability of such spacetimes, offer a window into the strong gravitational regime near compact objects. Following the approach of Refs.~\cite{Cunha_etal:2017PRL,Cunha_Herdeiro:2020PRL}, we identify LRs as critical points of the effective geodesic potentials \( H_{\pm}(\ell, \theta) \), defined in the orthogonal 2-space spanned by \( (\ell, \theta) \). These potentials, expressed in terms of the metric components, are:
	\begin{equation}
	    H(\ell,\theta)_{\pm}\equiv\frac{g_{t\varphi}\pm\sqrt{g_{t\varphi}^2-g_{tt}g_{\varphi\varphi}}}{g_{\varphi\varphi}}.	
	\end{equation}
Therefore, a LR is obtained whenever $\p_i H_{+}=0$ or $\p_i H_{-}=0$, or both conditions are met simultaneously, where $i=\ell, \theta$. In general, the sign of the $H(\ell,\theta)$ potential is associated with the rotation sense of the LR \cite{Cunha_Herdeiro:2020PRL}. 

It was established in Refs.~\cite{Cunha_etal:2017PRL,Cunha_Herdeiro:2020PRL} that a topological charge can be assigned to LRs by introducing a vector field ${\bm v}$ constructed from the normalized gradient of the $H(\ell,\theta)_\pm$:
	\begin{equation}
	{\bm v}=(v_\ell,v_\theta)=\left(\frac{\p_\ell H_\pm}{\sqrt{g_{\ell\ell}}},\frac{\p_\theta H_\pm}{\sqrt{g_{\theta\theta}}} \right)=\left(\p_\ell H_\pm, \frac{\p_\theta H_\pm}{\mathcal{R}}\right).
	\label{vcomponents}
	\end{equation}
Hence, LRs are characterized by the condition ${\bm v} = 0$. 

To assign a topological charge to LRs, we first delineate a contour, denoted as $\mathcal{C}$, within the $(\ell,\theta)$ plane. This contour is designed to be simple, closed, and piecewise smooth. We also introduce an auxiliary two-dimensional Cartesian space, represented by coordinates $(v_\ell, v_\theta)$ and referred to as $\mathcal{V}$. As we trace $\mathcal{C}$ in a positive direction---meaning in a counterclockwise manner---it maps onto a corresponding curve, $\mathcal{\tilde{C}}$, within $\mathcal{V}$. Due to the continuous nature of $\mathcal{C}$, the curve $\mathcal{\tilde{C}}$ similarly forms a closed path in $\mathcal{V}$. When the curve $\mathcal{C}$ encloses a LR in the physical 2-space $(\ell,\theta)$, the loop $\mathcal{\tilde{C}}$ circumscribes the origin $v=0$ in the auxiliary space $\mathcal{V}$.

In this auxiliary space $\mathcal{V}$, the vector field $\bm{v}$ can be expressed in polar coordinates:
\begin{align}
v_\ell &= v\cos\Omega, \\
v_\theta &= v\sin\Omega,
\end{align}
where $v$ represents the magnitude of the vector field $\bm{v}$, and $\Omega$ denotes the angle that $\bm{v}$ makes with the $v_\ell$ axis. 
	
By traversing a simple closed curve $\mathcal{C}$ that encircles a LR, we can assign a topological charge $w$ to the LR based on the total variation of the angle $\Omega$:
	\begin{equation}
	\oint_C d\Omega=2\pi w.
	\end{equation}
Each region bounded by the curve $\mathcal{C}$ contributes to the topological charge $w_\mathcal{C}$, with its value increasing or decreasing by precisely $\pm 1$. This ensures that the topological charge $w_\mathcal{C}$ remains constant as long as the enclosed number of LRs does not change. LRs with a topological charge of $w = -1$ are saddle points of the potential and called \textit{standard}, while those with $w = 1$ are extrema of the potential and called \textit{exotic}. According to the theorem in Refs.~\cite{Cunha_etal:2017PRL,Cunha_Herdeiro:2020PRL}, the topological charge derived from the winding number is $w=0$ for UCOs with trivial topology and $w=-1$ for asymptotically flat, stationary and axisymmetric BHs.

	\begin{figure}
	\centering
	\includegraphics[width=8.5cm]{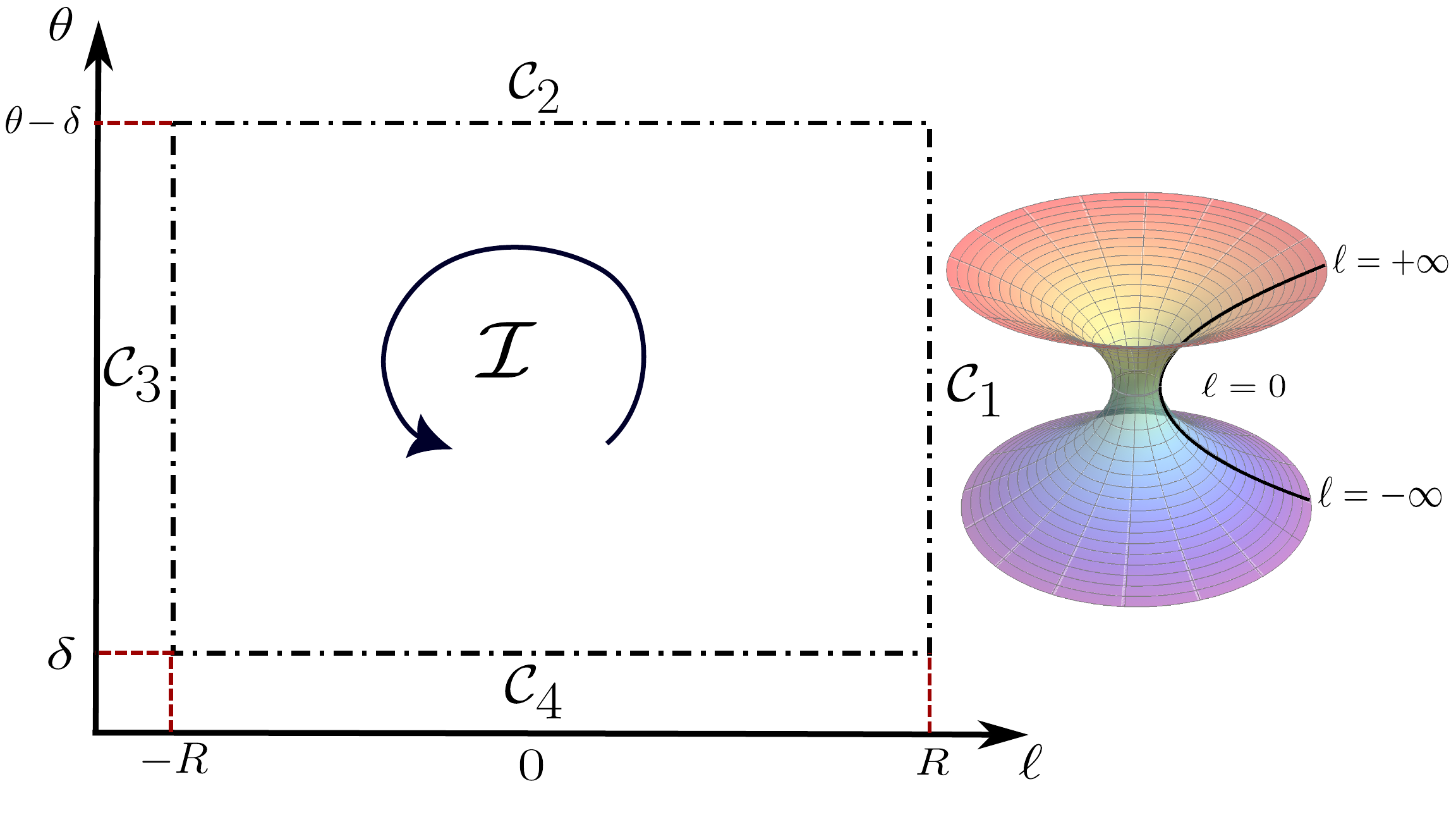}
	\caption{Diagram illustrating the integration contour $\mathcal{C}$, boundary of the region $\mathcal{I}$, which spans across both sides of the wormhole throat ($\ell=0$). The goal is to extend this integration curve comprehensively throughout the entire spacetime after taking the appropriate asymptotic limits.}
	\label{countour}
	\end{figure}
	 
	 We selected a contour $\mathcal{C}$ such that the wormhole throat is inside the subregion $\mathcal{I}$ enclosed by the curve $\mathcal{C}$, as illustrated in Fig.~\ref{countour}. The contour $\mathcal{C}$ is defined as:
	 \begin{equation}
	 \mathcal{C} = \bigcup_{i=1}^{i=4} \mathcal{C}_{i},
	 \end{equation}
	 with
	 	\begin{equation}
	 	\mathcal{C}_{i}=\begin{cases}
	 	i=1:\quad \ell=+ R,\ \delta\leq\theta\leq\pi-\delta,\\
	 	i=2:\quad \theta=\pi-\delta,\ -R\leq \ell\leq R, \\
	 	i=,3:\quad \ell=- R,\ \delta\leq\theta\leq\pi-\delta, \\
	 	i=4:\quad \theta=\delta,\ -R\leq \ell\leq R.
	 	\end{cases}
	 	\end{equation}
	 	
The topological charge \(w_\mathcal{C}\) can be computed, then, from the sum of the integrals along each curve:
	\begin{equation}
	2\pi w_\mathcal{C}=\sum_{i=1}^4 I_{i},
	\end{equation}
where	
	\begin{align}
	&I_1=\left[\int_\delta^{\pi-\delta}\frac{d\Omega}{d\theta}d\theta\right]_{r=R},\\
	&I_2=\left[\int_{R}^{-R}\frac{d\Omega}{d\ell}d\ell\right]_{\theta=\pi-\delta},\\
	&I_3=\left[\int_{\pi-\delta}^{\delta}\frac{d\Omega}{d\theta}d\theta\right]_{r=-R},\\
	&I_4=\left[\int_{-R}^{R}\frac{d\Omega}{d\ell}d\ell\right]_{\theta=\delta}.
	\end{align}
	
	The total topological charge for the wormhole spacetime, which connects two distinct universes, is derived by evaluating these integrals at both the asymptotic boundaries and the axis limits:
	\begin{equation}
	w=\lim_{R\to\infty}\left(\lim_{\delta\to 0}w_{\mathcal{C}}\right).
	\end{equation}

	\begin{itemize}
	    \item {\bf Axis limit:}
	\end{itemize}

The axis of symmetry is defined as the set of points where the Killing vector $\eta$ vanishes, which implies $g_{\varphi\varphi}=g_{t\varphi}=0$. In the axis limit, it is convenient to introduce the coordinate $\rho\equiv\sqrt{g_{\varphi\varphi}}$. As was discussed in Ref.~\cite{Cunha_Herdeiro:2020PRL}, at this limit, one can consider a small $\rho$ expansion obtaining that the potential $H_\pm$ behave as 
	\begin{equation}
		H_\pm\sim\pm\ \frac{\sqrt{-g_{tt}^0}}{\rho},
	\end{equation}
with $g_{tt}^0$ being the first term of the expansion of the metric component $g_{tt}$.

 From Eq.~\eqref{vcomponents} we can compute the $\theta$ component of the vector field $\bm{v}$, which will have the following behavior:
 \begin{equation}
v_{\theta} \sim\mp  \text{sign} \left( \frac{d\rho}{d\theta} \right) \frac{1}{\rho^2}.
\end{equation}

With this information, the angle $\Omega$ can be computed, yielding  
	\begin{equation}
	\Omega=\arcsin\left(\frac{v_\theta}{v}\right)\bigg\rvert_{0,\pi}\rightarrow\begin{cases}\pm\pi/2 , \hspace{0.1cm} \text{for}\hspace{0.1cm} \theta\rightarrow\pi,\\
	\mp\pi/2 , \hspace{0.1cm} \text{for}\hspace{0.1cm} \theta\rightarrow 0.\end{cases}
	\end{equation}

As in the BH case of Ref.~\cite{Cunha_Herdeiro:2020PRL}, the angle $\Omega$ is constant along the integration paths $\mathcal{C}_2$ and $\mathcal{C}_4$, which implies that $\{I_2, I_4\}$ have a null contribution to the topological charge $w_\mathcal{C}$.

        \begin{itemize}
            \item {\bf Asymptotic limits:}
        \end{itemize}
 
Using the asymptotic limit of the metric components given by Eqs.~\eqref{Nafconditions}, \eqref{afconditions} and \eqref{wafconditions}, we can infer the behavior of the potential $H_\pm$ when $R\to\infty$. At this limit, which corresponds to $\ell\to\pm\infty$, we have:
	\begin{equation}
		H_\pm(\ell,\theta)\sim\begin{cases}
		\pm\frac{1}{\ell\sin\theta},\ \text{when}\ \ell\to+\infty,\\
		\mp\frac{1}{\ell\sin\theta},\ \text{when}\ \ell\to-\infty.
		\end{cases}
	\end{equation}
	
	Consequently, one can conclude that the radial component of the vector $\bm{v}$ will have the following asymptotic behavior:
	\begin{equation}
	v_\ell\simeq\begin{cases}
	\ell\rightarrow+\infty, \quad\mp\frac{1}{\ell^2\sin\theta}\implies\text{sign}(v_\ell)|_\infty=\mp 1,\\
	\ell\rightarrow-\infty, \quad\pm\frac{1}{\ell^2\sin\theta}\implies\text{sign}(v_\ell)|_\infty=\pm 1.
	\end{cases}
	\label{vrinfty}
	\end{equation}

From Eq.~\eqref{vrinfty}, it is possible to establish that the vector $\bm{v}$ has a negative (positive) radial component along $\mathcal{C}_1$ for $H_{+}$ ($H_{-}$). As $\bm{v}$ approaches the intersection of the path $\mathcal{C}_1$ with $\mathcal{C}_2$, the direction of the vector field continuously shifts from inwards to upwards. Similarly, the direction of $\bm{v}$ changes from inwards to downwards as the vector field  approaches the intersection of $\mathcal{C}_1$ with $\mathcal{C}_4$.  This means that $\bm{v}$ winds in the negative direction along $\mathcal{C}_1$ when $\mathcal{C}$ is circulated in the positive direction, resulting in half of a full winding. Therefore:
	\begin{equation}
	\Omega^\infty_{\theta=\pi}-\Omega^\infty_{\theta=0}=-\pi.
	\end{equation}
	A similar reasoning can be applied to the path $\mathcal{C}_3$,	 yielding
		\begin{equation}
	\Omega^{-\infty}_{\theta=0}-\Omega^{-\infty}_{\theta=\pi}=-\pi.
	\end{equation} 
Hence, both asymptotic limits contributes by the same amount to the total topological charge.

        \begin{itemize}
            \item {\bf Total topological charge:}
        \end{itemize}
	Summing all the contributions in the appropriate limits, we obtain that the total topological charge for a stationary, axisymmetric, traversable and inter-universe wormhole is:
        \begin{align}
            w=\frac{1}{2\pi}\lim_{R\to\infty}\left[\lim_{\delta\to 0}\left(I_1+I_2+I_3+I_4\right)\right]=-1.
        \end{align}
It is important to remark that since we do not assume any evenness of the metric components with respect to the throat, our result is valid either to symmetric or asymmetric wormholes. Additionally, we note that the total topological charge is not affected by a smooth deformation of the throat. Therefore, our findings still hold even in cases of wormholes featuring a deformed throat with multiple bellies~\cite{Hoffmann:2017jfs,Hoffmann:2017vkf,Chew:2018vjp}.

The observation that wormholes yield similar outcomes to BHs suggests that they could potentially mimic some BH phenomenology. The unique topological characteristics of wormholes, however, may lead to observable signatures that are absent in BHs cases \cite{Paul_etal:2020,Olmo_etal:PLB2022,Guerrero_etal:PRD2022}.

 Additionally, the finding that these wormholes will always host at least one LR can be derived using the optical metric approach, particularly in the context of static and spherically symmetric cases, as detailed in Novo et al.~\cite{Cunha:2022nyw}. Further elaboration on this topic can be found in Appendix~\ref{appb}.
 
An illustration of the vector field $\bm{v}$ for some well-known wormhole solutions is presented in Appendix~\ref{appc}, where we can see how this vector field changes around standard and exotic LRs. We also show how the asymmetry with respect to the throat can affect the LR position.
 
\section{LR at the throat of the wormhole}\label{sec:3}
It can be demonstrated that for symmetric wormholes, a LR is always present at \( \ell = 0 \), that is, at the wormhole's throat. This result is due to the symmetrical nature of the geodesic potential \( H(\ell) \), ensuring that its derivative is equal to zero at the origin \( \ell = 0 \).  In the discussion that follows, we provide a proof of this assertion.

Let \( H(\ell) \) be a continuous and differentiable even function defined on the interval \((-\infty, +\infty)\). Since \( H(\ell) \) is an even function, it satisfies the property \( H(\ell) = H(-\ell) \) for all \( \ell \) within its domain. The even nature of \( H(\ell) \) implies that its derivative \( H'(\ell) \) is an odd function, fulfilling the condition \( H'(-\ell) = -H'(\ell) \). Consequently, at \( \ell = 0 \), we have:
\begin{equation}
    \frac{d H}{d \ell}\bigg\rvert_{\ell=0} = -    \frac{d H}{d \ell}\bigg\rvert_{\ell=0}.
    \label{dhzero}
\end{equation}
Equation~\eqref{dhzero} holds true only if \( H'(0) = 0 \). Consequently, \( H(\ell) \) has a critical point at \( \ell = 0 \), the midpoint of the interval \((-\infty, +\infty)\). From these observations, it follows that {\it for any stationary, axisymmetric, asymptotically flat wormhole, symmetric with respect to the throat, there always exists one LR located at the throat.}  A similar conclusion for the static case has been previously discussed in Ref.~\cite{Bronnikov:2018nub}.

Based on the contrapositive of the previously stated proposition, one can infer that a lack of a LR at the throat suggests asymmetry in the wormhole's structure relative to the throat. However, it is essential to acknowledge that the presence of a LR at the throat in asymmetric wormholes is still possible -- asymmetry does not necessarily prevent the existence of a LR at the throat.

Employing a similar reasoning, it can be demonstrated that in any stationary, axisymmetric single BH spacetime with $\mathbb{Z}_2$ symmetry~\footnote{Spacetimes whose metric components are symmetric with respect to the equatorial plane.} there is invariably a LR located on the equatorial plane. This finding was recently demonstrated in Ref.~\cite{Zeus_etal:2024}.

\section{Final remarks}\label{sec:remarks}
In order to test the Kerr hypothesis, several alternatives to BH foils were proposed in the literature. The current and future observational evidence from gravitational waves and shadow images is expected to provide continuous data to investigate the nature of astrophysical compact objects. It is well understood that the shadow cast by astrophysical compact objects, a key observable feature in these investigations, is intimately connected to the properties of LRs. 

Given this intrinsic connection, our work generalized the LR theorems of Ref.~\cite{Cunha_etal:2017PRL} to the case of UCOs with non-trivial topology. By applying the technique developed in Ref.~\cite{Cunha_Herdeiro:2020PRL} to a general class of inter-universes wormholes, we demonstrated that these entities will always have at least one standard LR.

We have also shown that wormholes symmetric concerning the throat will always present a LR at the throat. This is a consequence of the even symmetry of the effective null geodesic potential $H(\ell,\theta)_\pm$.

We would like to remark that although wormholes are in the same topological class as BHs (concerning LR physics), the former lacks a known dynamic formation mechanism. Topological non-triviality requires new physics, since gravitational collapse within GR is not expected to change the topology of spacelike sections unless causality is violated \cite{Geroch:1967fs}.
	
Exploring intra-universe wormholes, which could theoretically connect different locations within the same asymptotic region, presents an interesting and natural follow-up of our work. However, the absence of exact solutions for such objects marks a significant challenge. Although local physics near the throat is independent of the topology of the spacetime, which means that observers close to the throat cannot distinguish if they are in an inter-universe or in an intra-universe wormhole, the multi-connectedness characteristics of such geometry will have an impact on the global properties of the spacetime. Therefore, the LR structure could be different.	

\begin{acknowledgments}
	
	We would like to thank João P. A. Novo and Pedro V. P. Cunha for their comments during the writing of this paper. We are grateful to Funda\c{c}\~ao Amaz\^onia de Amparo a Estudos e Pesquisas (FAPESPA), Conselho Nacional de Desenvolvimento Cient\'ifico e Tecnol\'ogico (CNPq) and Coordena\c{c}\~ao de Aperfei\c{c}oamento de Pessoal de N\'ivel Superior (CAPES) -- Finance Code 001, from Brazil, for partial financial support. SX and LC thank the University of Aveiro, in Portugal, for the kind hospitality during the completion of this work.	This work is supported  by the  Center for Research and Development in Mathematics and Applications (CIDMA) through the Portuguese Foundation for Science and Technology (FCT -- Fundaç\~ao para a Ci\^encia e a Tecnologia), https://doi.org/10.54499/UIDB/04106/2020
and https://doi.org/10.54499/UIDP/04106/2020. The authors would also like acknowledge support from the projects
http://doi.org/10.54499/PTDC/FISAST/3041/2020, as well as http://doi.org/10.54499/CERN/FIS-
PAR/0024/2021 and https://doi.org/10.54499/2022.04560.PTDC.  This work has further been supported by  the  European Horizon Europe staff exchange (SE) programme HORIZON-MSCA-2021-SE-01 Grant No.~NewFunFiCO-101086251.
\end{acknowledgments}

\appendix
\section{Light rings and the optical metric\label{appb}}

Recent studies have used the optical metric --- a two-dimensional Riemannian metric perceived by massless particles --- to investigate LRs in BH spacetimes~\cite{Cunha:2022nyw, Qiao:2022jlu}. These works effectively recover and extend the findings of Refs.~\cite{Cunha_etal:2017PRL,Cunha_Herdeiro:2020PRL}. In this Appendix, we demonstrate that the optical metric method is equally applicable to asymptotically flat, static, and spherically symmetric wormholes, as described by the line element
\begin{equation}
ds^2 = -f(\ell)dt^2 + h(\ell) d\ell^2 + \mathcal{R}(\ell)^2 d\Omega^2,
\label{eqb1}
\end{equation}
where $d\Omega^2$ denotes the metric on the unit 2-sphere, and the functions $f(\ell)$, $h(\ell)$, and $\mathcal{R}(\ell)$ are guaranteed to be at least $C^2$ smooth.

From Eq.~\eqref{eqb1}, the optical metric is derived by setting $ds^2=0$ and solving for $dt$, which yields
 \begin{equation}
     dt^2 = \frac{1}{f(\ell)} \left( h(\ell)d\ell^2  + \mathcal{R}(\ell)^2 d\phi^2 \right).
    \label{eqb2}
 \end{equation}
We restricted our analysis to the equatorial plane $\theta=\pi/2$, without any loss of generality. 
 
In this approach, LRs are identified as the roots of the geodesic curvature $\kappa_g$ of circular orbits:
        \begin{equation}
	       \kappa_g = \sqrt{\frac{h(\ell )}{f(\ell )}}\frac{\left(2 f(\ell ) \mathcal{R}'(\ell )-\mathcal{R}(\ell ) f'(\ell )\right)}{2 \mathcal{R}(\ell )}.
	      \label{eqb3}
    \end{equation}

Given that LRs satisfy $\kappa_g=0$, the number of LRs can be determined by examining the asymptotic behavior of Eq.~\eqref{eqb3}. The condition of asymptotic flatness \eqref{afconditions} implies:
    \begin{equation}
        \lim_{\ell\to\pm\infty}\kappa_g(\ell)=\lim_{\ell\to\pm\infty}\frac{\mathcal{R}'(\ell )}{\mathcal{R}(\ell)}\sim\begin{cases}
        -\frac{1}{\ell} \quad \text{as}\ \ell\to-\infty,\\
        \frac{1}{\ell} \quad \text{as}\ \ell\to+\infty,
        \end{cases}
    \end{equation}
indicating that $\kappa_g$ approaches zero asymptotically from positive (negative) values, as $\ell$ goes to $+\infty$ ($-\infty$). This follows directly from Eq.~\eqref{afconditions}.

The intermediate value theorem asserts that if a continuous function changes sign over an interval $[a, b]$, there must be at least one point within that interval where the function vanishes. Applying this principle, the sign change of the geodesic curvature $\kappa_g(\ell)$ from negative to positive, as $\ell$ moves from $-\infty$ to $+\infty$, requires that $\kappa_g(\ell)$ intersects the $\ell$-axis at least once, indicating the presence of an odd number of LRs.

It is important to note that this analysis does not rely on any symmetry regarding the wormhole's throat, making the result applicable across a wide range of wormhole geometries, whether symmetric or asymmetric.

\section{Vector field of traversable wormholes}\label{appc}

	In this Appendix, we exhibit the behavior of the vector field $\bm{v}$ for two different wormhole solutions. The chosen solutions  exemplify spacetimes that are symmetric or asymmetric with respect to the throat. The symmetric case presents multiple LRs, and the total topological charge does not change.

\subsection{Symmetric wormhole: Simpson-Visser wormhole}
	\begin{figure}
	\centering
	\includegraphics[width=7.5cm]{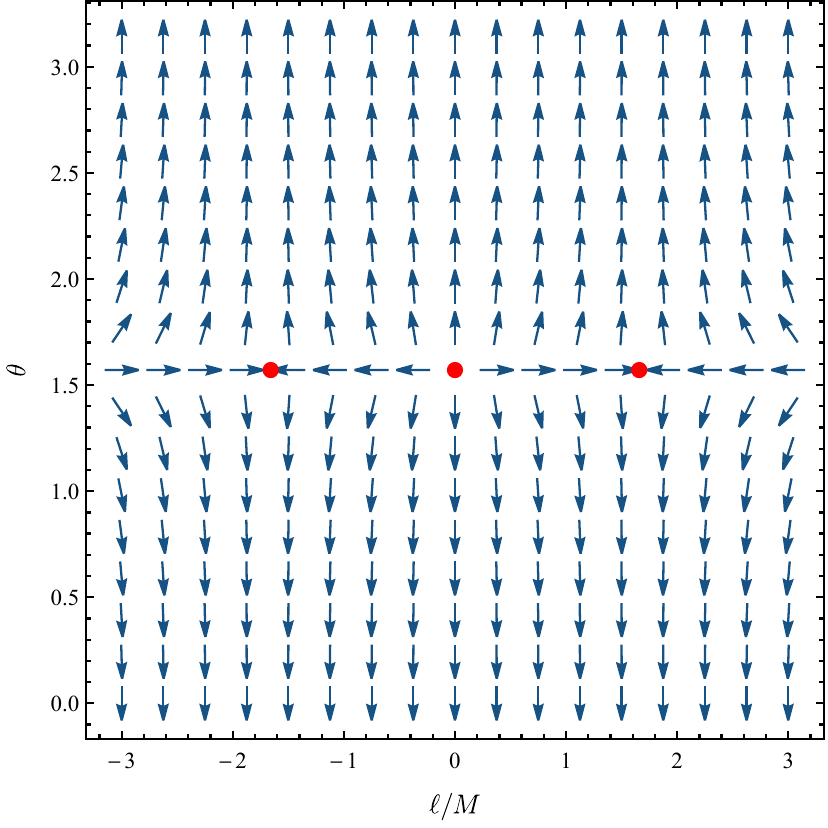}
	\caption{Vector field $\bm{v}$ for the Simpson-Visser wormhole with $a=2.5M$. The red dots represent the locations of the LRs.}
	\label{vectorSV}
	\end{figure}
The Simpson-Visser solution~\cite{Simpson:2018tsi} is a static and spherically symmetric spacetime given by Eq.~\eqref{eqb1}, where the metric functions are
	\begin{align}
	&f(\ell)=h(\ell)^{-1}=\left(1-\frac{2 M}{\sqrt{\ell^2+a^2}}\right),\\
	&\mathcal{R}(\ell)^2=\ell^2+a^2.
	\end{align}

When $a>2M$, we have a traversable symmetric wormhole geometry. The effective geodesic potentials $H_\pm(\ell,\theta)$ for this spacetime are given by
	\begin{equation}
		H_\pm(\ell,\theta)=\pm \frac{\sqrt{\left(1-\frac{2M}{\sqrt{\ell^2+a^2}}\right)(\ell^2+a^2)}}{\left(a^2+\ell^2\right)\sin\theta},
	\end{equation}
	and the components of the vector field $\bm{v}$ are
	\begin{align}
		&v_\ell=\mp\frac{\ell  \csc \theta  \left(\sqrt{a^2+\ell ^2}-3 M\right)}{\left(a^2+\ell ^2\right)^2},\\
		&v_\theta=\mp\cot\theta\csc \theta \frac{\sqrt{1-\frac{2 M}{\sqrt{\ell^2+a^2}}}}{\left(a^2+\ell ^2\right)^{3/2}}.
	\end{align}
	
	Depending of the value of parameter $a$, this geometry can present more than one LR. For instance, Fig.~\ref{vectorSV} exhibit the case for $a=2.5M$, where we notice the presence of two unstable LRs on each side of the throat and one stable LR at the throat. 	Note that the behavior of the vector field around the stable LR changes, indicating that this LR has a different value of topological charge ($w_{\text{stable}~\sss\text{LR}}=+1$). Nevertheless, the {\it total} topological charge of the spacetime remains unchanged.
\subsection{Asymmetric wormhole: Ellis-Bronnikov wormhole}
\begin{figure}
	\centering
	\includegraphics[width=7.5cm]{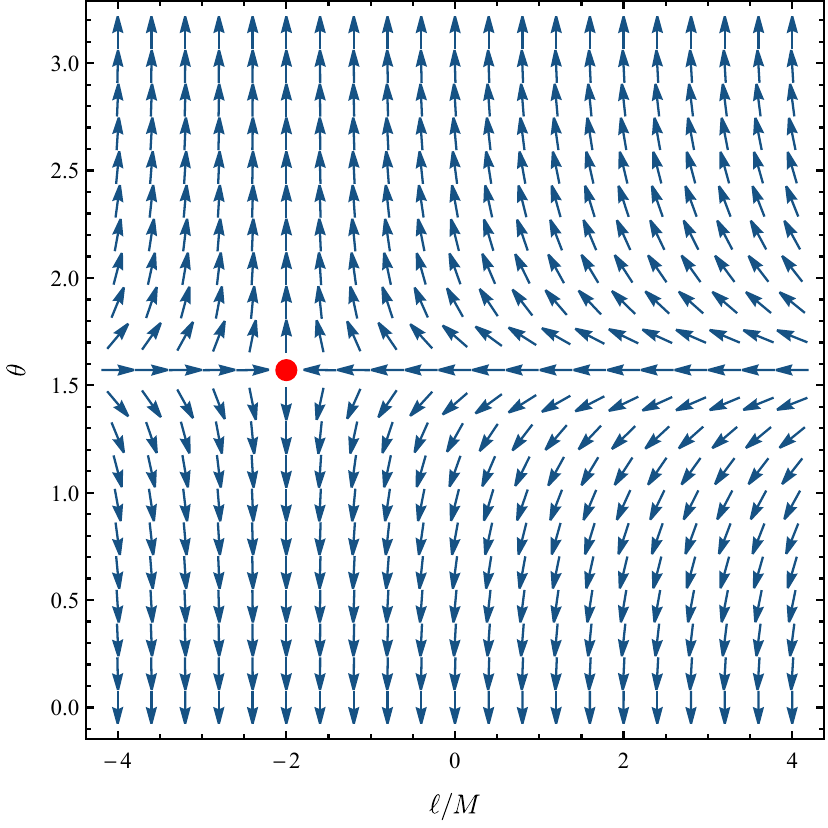}
	\caption{Vector field $\bm{v}$ for the Ellis-Bronnikov wormhole with $q=4.4M$. The red dot shows the location of the LR.}
	\label{vectorEB}
	\end{figure}
	The Ellis-Bronnikov spacetime is a static and spherically symmetric solution of GR coupled to a free phantom scalar field~\cite{Huang_etal:2023,Bronnikov:1973fh,Yazadjiev:2017twg}. The metric functions are
	\begin{align}
		&f(\ell)=h(\ell)^{-1} = e^{-\frac{M}{q} \Phi(\ell)}, \quad \mathcal{R}(\ell)^2 = \frac{\ell^2 + q^2 - M^2}{h(\ell)},\\
		&\Phi(\ell) = \frac{2q}{\sqrt{q^2 - M^2}} \arctan\left(\frac{\ell}{\sqrt{q^2 - M^2}}\right).
	\end{align}
	This solution is described by two parameters: the mass $M$ and the scalar charge $q$. When $M=0$, the solution reduces to the well-known Ellis wormhole~\cite{Ellis:1973yv}. For $M\neq 0$, the Ellis-Bronnikov solution describes an asymmetric wormhole spacetime, which is our focus here.
	
	The effective geodesic potentials $H_\pm(\ell,\theta)$ for this geometry are given by
	\begin{equation}
		H_\pm(\ell,\theta)=\pm \frac{h(\ell)}{\sin\theta\sqrt{\ell^2+q^2-M^2}},
	\end{equation}
	and the components of the vector field $\bm{v}$ for the Ellis-Bronnikov wormhole are
	\begin{align}
		&v_\ell=\mp\frac{\csc \theta ~(2m+\ell)\sqrt{h(\ell)}}{(\ell^2+q^2-m^2)^{3/2}},\\
		&v_\theta=\mp\frac{\cot\theta\csc \theta~(2m+\ell)~\sqrt{h(\ell)}}{(\ell^2+q^2-m^2)^{3/2}}.
	\end{align}
	
From Fig.~\ref{vectorEB}, we observe the presence of only one unstable LR, which lies outside the throat in one of the universes. This happens due to the asymmetry nature of the Ellis-Bronnikov spacetime. 
\color{black}
	{}
\end{document}